# Spin dependent transport and recombination in organic light-emitting diodes

George B. Silva and Carlos F.O. Graeff[a]

Departamento de Física e Matemática, FFCLRP-USP, Av. Bandeirantes 3900, 14040-901

Ribeirão Preto, Brazil

Frank Nüesch and Libero Zuppiroli

LOMM-IMX-EPFL, CH-1015 Lausanne, Switzerland

Electrically Detected Magnetic Resonance (EDMR) was used to study a series of multilayer organic devices based on aluminum (III) 8-hydroxyquinoline. These devices were designed to identify the microscopic origin of different spin dependent process, i.e. hopping and exciton formation. EDMR is demonstrated to probe molecular orbitals of charge carriers, and thus indirectly explore interfaces, exciton formation, charge accumulation and electric fields in operating organic based devices.

71.35.-y, 73.40.-c, 73.61.Ph, 76.90.+d, 85.60.Jb, 85.65.+h

---

[a] corresponding author, e-mail: cfograeff@ffclrp.usp.br

Since the demonstration of highly efficient, high brightness organic light emitting devices (OLED) achieved by using a multilayer structure, there is a growing interest for detailed understanding of electro-optical properties of these devices and their constituent materials[1]. Despite this, aluminum (III) 8-hydroxyquinoline ($Alq_3$) though the prototype molecular emitter still has many of its structural and electro-optical properties inadequately understood in detail. The efficiency of an OLED is determined by the fraction of injected electrons and holes that recombine to form emissive states. This is a multifaceted problem that involves balanced charge injection, carrier mobility among different layers, charge distribution at interfaces, and exciton formation. Unfortunately there are few experimental techniques that can address these questions under device operation conditions. Electron Spin Resonance (ESR), a spectroscopy that can directly measure the spin density distribution, in other words, that can probe the wavefunctions at and surrounding unpaired spin sites, has had little success with the molecules used on OLEDs[2]. The difficulty is that these molecules are in general diamagnetic and becomes paramagnetic, or ESR active, only when an electron is injected or extracted from its molecular orbital, in other words when they are singly charged. To generate ESR active molecules is not simple, for example it can involve doping using reducing or oxidizing agents or illumination. In the latter, contamination may be an issue, and actually dominate the ESR response. In contrast, non-conventional electron magnetic resonance (EMR) techniques, such as electrically detected magnetic resonance (EDMR) allow the direct study of electron spins in diamagnetic systems, as will become clearer in the following. EDMR though not a new technique has gained a new impulse in recent years due to its superior sensitivity with respect to ESR, among other advantages.[3] As a matter of fact EDMR is one of the few EMR spectroscopies that is applicable to nanotechnology.[3] Typically, in an EDMR experiment spin level transitions induced by magnetic resonance are measured through changes in device current. The key to EDMR is that many of the processes that lead to charge transport and recombination are spin dependent. In this work charge injection, transport and recombination (exciton formation) in operating $Alq_3$ based devices are investigated using EDMR.

All devices used in this work were fabricated by thermal evaporation in high vacuum (< 5 × $10^{-7}$ mbar). Details about sample structure are given in table 1. In addition to the multilayered state-of-the art OLEDs, simpler unipolar devices were fabricated. These specially designed devices were used primarily to allow identification of spin dependent processes in OLEDs. A more detailed description on similar diodes concerning deposition, structure and electro-optical characteristics is given elsewhere.[4] To avoid air induced degradation, the devices were directly transferred to an inert gas glove box after fabrication without being exposed to the atmosphere. Samples with lateral dimensions no greater than 3 mm were then contacted and sealed inside an ESR quartz tube. EDMR was measured using modified commercial X-Band and K-band spectrometers under different bias conditions.[3]

Typical EDMR spectra for different devices are shown as full lines in Figure 1. To better visualize the sign of the current change, the spectra in this figure have been integrated, see fig. 4 for a typical non processed (not integrated) signal. As can be seen from this figure and table 1, EDMR in unipolar devices and bipolar OLEDs have distinct characteristics; for example, opposite sign and very different amplitudes (x25). As will be discussed, these differences reflect the distinct spin dependent mechanism present in unipolar devices as opposed to bipolar devices. As expected, all devices listed in Table 1 are ESR inactive.[2] Let us start with the unipolar devices, whose spin dependent mechanism is well known in several inorganic as well as organic semiconductors.[3] In electron-only (e-only) devices, an EDMR signal is observed for all samples having a LiF interface layer at the aluminum electrode with a thickness >0.8 nm. Furthermore, the g-factor and peak-to-peak linewidth ($\Delta H_{pp}$) are dependent on the applied bias. EDMR in e-only devices is assigned to the formation of dianions from the hopping of an electron (or spin) in neighboring anions. To understand this spin dependent mechanism in more detail, the process will be described as a charge transfer (CT) reaction. This CT reaction is spin dependent since the precursors are two spins forming a pair (neighboring anions) while the product is a molecule with two anti-paralel spins (dianions). The precursor spin pair can either have the spins parallel or anti-parallel, however spin pairs with parallel spins will not react, due to spin conservation rules. As a consequence, under steady state conditions, there is an excess of precursors with parallel spins. In resonance spins are flipped or in other words, parallel spins are converted to antiparallel spins whose CT reaction is allowed, and thus the device current increases. Doubly charged molecules (dianions) are not believed to be formed in the bulk of $Alq_3$ but rather at interfaces, where charge accumulation occurs[1]. Charge accumulation is enhanced by using an insulating LiF layer that acts as a barrier for carrier transfer from $Alq_3$ to the positive Al electrode.[5] It is exactly this interface that we believe EDMR is probing, and explains why EDMR is so sensitive to LiF layer thickness. We have also investigated hole-only (h-only) devices using a single layer device based on the hole conductor (N, N'-di(naphthalene-1-y1)-N, N'-diphenyl-benzidine) ($\alpha$-NPD). The EDMR signal in this device is similar to the one found in the e-only device, as shown in Table 1. Unlike e-only devices, for h-only devices $\Delta H_{pp}$ is narrower and the signal is only weakly bias dependent. Analogously it is believed that the spin dependence in h-only devices is related to charge hopping close to the Ag electrode, as a result of the bad injection characteristics of this interface[1].

We can now analyze EDMR in bipolar multilayer devices commonly used in standard OLEDs. The EDMR signal can only be observed when the diode starts to luminesce, or in other words, when excitons are formed. Furthermore, the signal has two distinct components (Figure 1). To demonstrate that the signal has two components we will make use of a peculiar characteristic of EDMR not commonly found in other EMR spectroscopies. Dersch et al.[6] have shown that differences in spin dependent relaxation times can be used to separate different contributions (components) of the EDMR signal. In our case, the two components have different g-factors (see Table 1). So if two EDMR components have different relaxation times, one expects different phase shifts with respect to the Lock-in reference signal, as shown in Figure 2. The two channels of the Lock-in amplifier can be represented as orthogonal axes and the signal as a vector (see insets to Figure 2). In this representation, the signal measured in one of the two channels, is the projection of the signal vector. Note that as represented in Figure 2, under appropriate conditions the two components can be isolated. For example, in the inset on the upper left, the component (e) is perpendicular to channel 1 while the component (h) is not. Thus in this condition channel 1 is only measuring the h-component. This two component signal found exclusively in bipolar OLED is assigned to spin dependent exciton formation. The phase shifted components of the signal are attributed one to electrons and the other to holes. Hopping related signals are also observed in OLEDs, yet as seen in Figure 1 at room temperature, the hopping related signal is probably masked since the exciton related signal is at least one order of magnitude stronger. At low temperatures T < 200K, where spin dependent hopping is enhanced due to additional space charge, the EDMR signal is composed of both hopping and exciton related signals.

Spin dependent exciton formation has a mechanism similar to the one described earlier related to charge hopping. However the CT reaction product is in the excited state, and thus can be either a singlet or a triplet (figure 3). To understand the microscopic mechanism behind spin dependent exciton formation, we will adapt the description of Wohlgenannt et al.[7] originally proposed for Optically Detected Magnetic Resonance to EDMR. Note that though this spin dependence is well established in molecular systems,[7] it has not so far been identified when using EDMR.[3] In this case the changes in current will be determined, not by the CT step as in the hopping case, but rather by the lifetime of the product. Triplets are long lived species when compared to singlets, so the former contribute less to the overall device current. For example, if in resonance the generation of triplets is enhanced a decrease in current is expected. The key to Wohlgenannt's model is that it is assumed that spin-dependent exciton formation cross-section for singlets ($\sigma_S$) and triplets ($\sigma_T$) are not necessarily the same. As seen on the left side of figure 3, the exciton precursors are a neighboring spin pair formed by a hole (or cation) and an electron (or anion). Parallel spin precursors form only triplet states ($T_{+1}$ and $T_{-1}$), thus in this case the reaction rate ($R_P$) is proportional to $2\sigma_T$. Using similar arguments the reaction rate between antiparallel spin pairs ($R_{AP}$) is proportional to ($\sigma_S + \sigma_T$). The proportionality constant is assumed to be the same in both cases. The sign of the current change is determined by $\sigma_S/\sigma_T$. If $\sigma_S < \sigma_T$ then $R_{AP} < R_P$, and as a consequence, there is an excess of antiparallel spin pairs when out of resonance. Under saturated magnetic resonance conditions the pair densities with parallel and antiparallel spins become equal. In other words, antiparallel pairs are partially converted into parallel pairs. As a result, there is a change in the relative generation rate of the CT reaction products. In resonance, the generation of long lived triplet excitons increases, which induces a decrease in device current at fixed bias. Thus, our results indicate that surprisingly in Alq$_3$ based OLEDs $\sigma_S < \sigma_T$. $\sigma_S/\sigma_T$ ratio has been the subject of numerous investigations lately as a result of its importance in what concerns the OLED efficiency. Interestingly in the discussion so far it is implicit that $\sigma_S \geq \sigma_T$ [7,8]. In fact it is found empirically that the larger the molecule/polymer the larger is $\sigma_S/\sigma_T$. On the other hand, our finding is in good agreement with previous reported work, which showed that in Alq$_3$ based OLEDs the singlet fraction is (22 ± 3) %, a value close but smaller than 25% when $\sigma_S = \sigma_T$ [9].

As discussed in the introduction one of the interesting features of EMR is that it is a probe of the spin/electron wavefunction and its neighborhood. In our case, to be more specific, this information is encoded in the g-factor and lineshape. Before discussing in detail the origin of g-factors and lineshapes found in Alq$_3$ based devices, we would like to stress that in some cases, EDMR probes two distinct spin systems simultaneously. In the exciton related signal for example, the microscopic information about the hole (or electron) molecular orbital and surroundings is available only if the hole (or electron) component can be isolated. This may in fact be exclusive to carbon based systems, such as $Alq_3$. The reason lies in the stronger spin-orbit coupling found in heavier atoms, which constitute inorganic materials. Strong spin orbit coupling implies for example short spin relaxation times, in other words phase lags between precursor pairs that are very difficult to be separated. For example, in Dersch's work [6] it was not the recombining pair components that were separated, but rather signals coming from distinct spin dependent mechanisms. The lineshape of EMR signals coming from spins occupying π orbitals are known to be dominated by hyperfine coupling to protons (H).[10] This has an interesting consequence; molecular orbitals that are more localized in nature are expected to have smaller $\Delta H_{pp}$. In our materials the orbitals implied in electron transport are called lowest occupied molecular orbital (LUMO), while those related to hole transport are named highest occupied molecular orbital (HOMO). As seen in Table 1, our results indicate that the $Alq_3$ LUMO is more delocalized than the HOMO in $Alq_3$ and α-NPD, which is in good agreement with reports in the literature.[11] On the other hand, in these organic molecules the g-factor can be used as a probe of the atoms which are close to the molecular orbital.[12] Heavier atoms have greater spin-orbit coupling which leads to greater g-factor deviations from the free-electron case (2.0023). From our experiments, we conclude that the HOMO lies closer to the oxygen atom in the ligand of $Alq_3$ than does the LUMO, in accordance with previous works.[11] Analogously, the molecular structure of α-NPD has no oxygen, thus as expected, the HOMO related signal has a g factor closer to 2. The fact that EDMR can distinguish a hole in α-NPD or $Alq_3$ through the g-factor has an important consequence concerning exciton formation dynamics. Our data show that excitons are formed from $Alq_3$ holes, in other words, holes must be injected in the $Alq_3$ layer before excitons are formed. In fact, EDMR measurements in OLEDs where α-NPD is absent have a signal identical to the one shown in Figure 1.

Since EDMR is an EMR technique allowing the measurements of spin states in operating devices, effects due to strong internal electric fields have to be considered. These fields not only originate from the applied electric field, but also from internal charge accumulation. Electric field effects are known to influence the ESR response[12-14]. These effects can be visualized as distortions of the electronic charge cloud induced by the E-field. In the case of our devices the E-field has a well defined direction; perpendicular to the substrate. In this configuration the distortions in the electronic cloud generate anisotropic g-shifts, as seen in Figure 4. The shifts are observed in both bipolar and homopolar devices and are further evidence that EDMR is probing processes close to interfaces, where E-fileds are particularly strong[4]. The g-shift has axial symmetry as anticipated, and more important the magnitude and sign of the shifts are compatible with the ones predicted by theory[12-14]. As expected the more delocalized LUMO has greater g-factor anisotropy, as well as a stronger bias dependence than those at the HOMO. We would like to point out that the bias dependence of the EDMR signal is probably dominated by a change in charge density close to the precursor spin pair. For example, in the case of the exciton related signal, the process occurs close to the $Alq_3$/α-NPD interface, where the device architecture constrains high accumulation of electrons and, at high bias, also of holes[1,4]. Thus it is expected that an increase in bias, which is followed by an increase in charge (or spin) density close to the precursor spin pair, will induce a

$\Delta H_{pp}$ broadening through exchange interaction. The g-factor dependence on bias is also attributed to local charges close to the spins. These randomly distributed local charges will generate E-fields that cause randomly induced distortions of the electronic cloud and are responsible for the g-factor dependence on bias. Understanding these dependencies would require a detailed understanding of charge distribution, spin dependent charge transport and recombination that we do not have, especially in a multilayered structure such as our OLED[1,4]. Note that in the EDMR literature the E-field effect has hardly been discussed.[3] This is not surprising since E-field effects in EMR are not expected in inorganic semiconductors[14], and especially in tetrahedral coordinated atoms such as Si.

This work was supported by the FNSF (Switzerland) contract No. 20-67929.02, as well as the following Brazilian agencies: FAPESP, CNPq, CAPES and IMMP/MCT. The authors acknowledge Philippe Bugnon for technical support.

Table Captions

**Details of devices used and EDMR signal characteristics. Notice that the OLED (bipolar) is the only device with a signal with two components. The variations in EDMR signal characteristics shown are relative to the applied bias. The numbers inside parenthesis are the errors associated to the measured quantity. $g_\parallel$ is the g-factor when the external magnetic field direction is parallel to the direction of the device internal electric field. $g_\perp$ when the same fields are perpendicular. The values displayed for the g-factor and linewidth ($\Delta H_{pp}$) correspond to parallel magnetic and electric fields directions.**

| Device | Structure | EDMR comp. | g-factor (g-2) × $10^4$ | $\Delta H_{pp}$ (mT) | Bias depend. | $g_\parallel - g_\perp$ (× $10^4$) |
|---|---|---|---|---|---|---|
| OLED (bipolar) | ITO/CuPc(12nm)/ α-NPD(40nm)/Alq$_3$(60nm)/ LiF(0.8nm)/Al | electrons | 27 – 37(3) | 2.0 – 3.4(0.3) | strong | 6(2) |
| | | holes | 42(3) | 1.5(0.3) | weak | 6(2) |
| e-only | Al/LiF(x)/Alq$_3$(150nm)/LiF(x)/Al 0< x < 8 nm | ----- | 26-31(3) | 2.0-2.6(0.3) | strong | 10(5) |
| h-only | ITO/α-NPD(180nm)/Ag | ----- | 31(5) | 1.5(0.5) | weak | 6(5) |

Figure captions

**Figure 1:** Typical EDMR signals found in Alq$_3$ based devices are shown as full lines. Notice that the signal in the e-only device is much weaker than the signal found in the bipolar OLEDs, and opposite in sign. The OLED signal is composed of two Gaussians that are plotted as dashed lines. The sum of the two Gaussians (fit) is also shown as a dotted line.

**Figure 2:** g-factor as a function of reference signal phase for an EDMR signal coming from an OLED, measured in K-band. Notice the abrupt change in phase close to 90˚. The inset to the figure shows the signal component vectors and their projection in the Lock-in channels, details are given in the text.

**Figure 3:** Schematic diagram of the spin dependent exciton formation. On the left the precursor spins are represented, while on the right the charge transfer reaction product, an exciton as well as a neutral molecule. The spin configurations on the right side are representative only of the exciton.

**Figure 4:** Typical EDMR signals for an OLED in two different orientations with respect to the external magnetic field. The electric field is perpendicular to the device surface, or in other words parallel to the growth direction. Notice that when the diode has the electric field direction parallel to the external magnetic field, the g-factor increases.

Figure 1-4                    Barbosa et al.

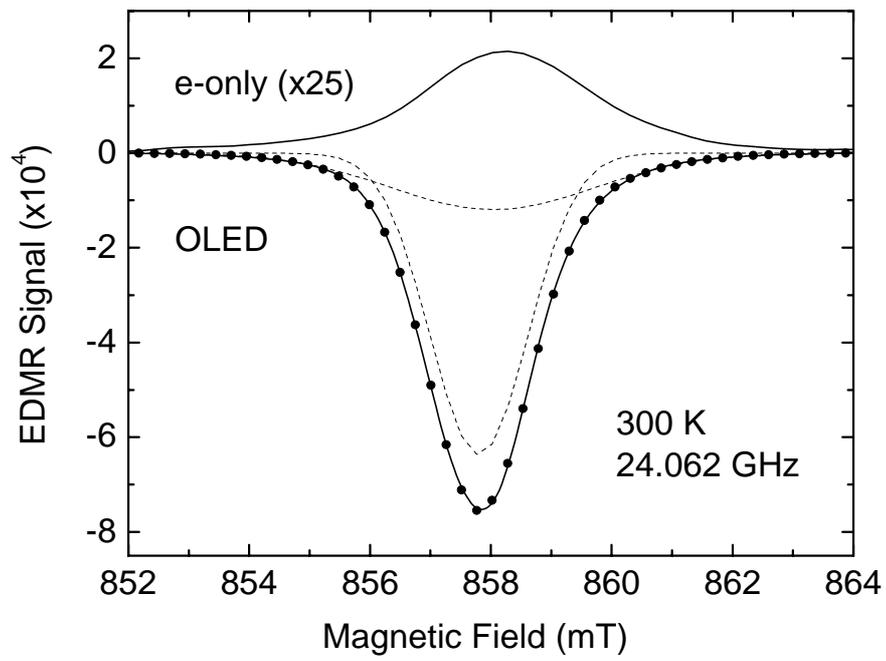

Figure 2-4    Barbosa et al.

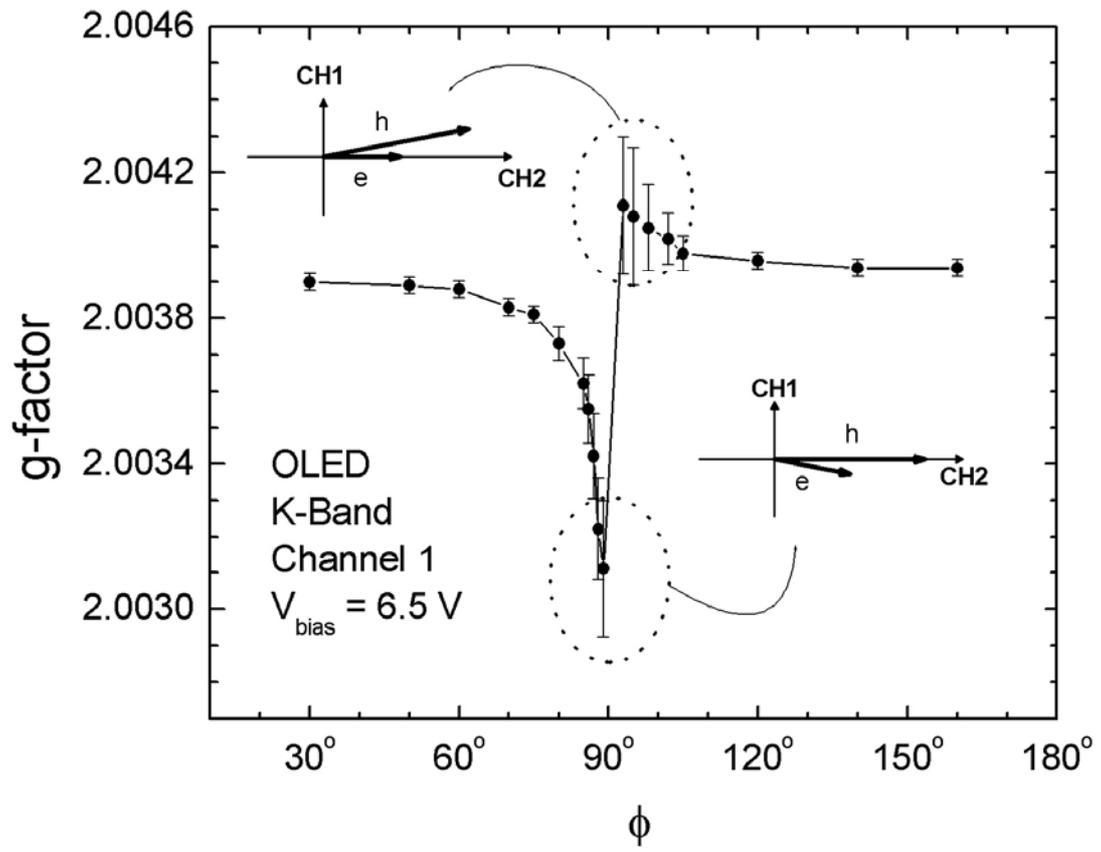

Figure 3-4                Barbosa et al.

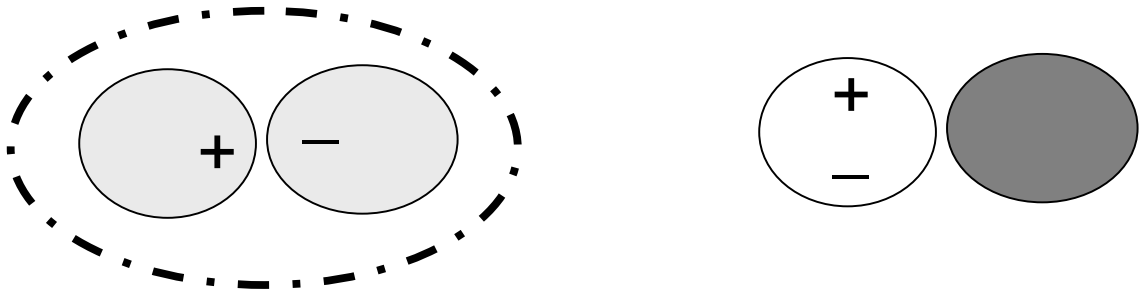

**Before**                              **After**

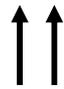   $R_P$
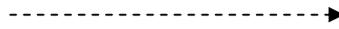   $T_+$ or $T_-$

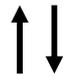   $R_{AP}$
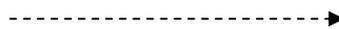   $S_1$ or $T_0$

Figure 4-4                    Barbosa et al.

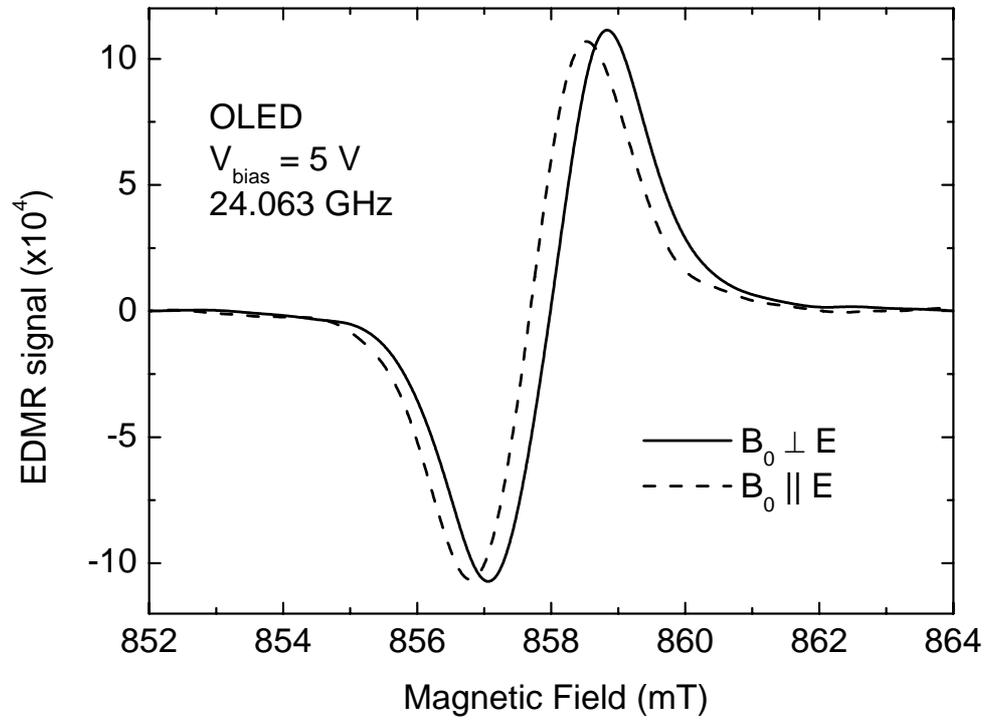